\documentclass[aps,prb,reprint]{revtex4-2}
\usepackage{graphicx} 
\usepackage{amsmath} 
\usepackage{bm} 
\usepackage{natbib}
\usepackage{natmove}
\usepackage[colorlinks=true, linkcolor=blue, citecolor=blue, urlcolor=blue, linktoc=page, bookmarks=false, pdfstartview={FitH}, pdfborder={0 0 0.0 [3 3]}]{hyperref} 
\usepackage{color} 
\usepackage{dcolumn} 
\newcolumntype{.}{D{.}{.}{0.3}}
\newcolumntype{-}{D{.}{.}{4.0}}
\usepackage{makecell}
\usepackage{multirow}
\usepackage{cleveref}
\usepackage{easyReview} 
\crefname{figure}{Fig.}{Figs}
\crefname{table}{Table}{Tables}
\crefname{section}{Sec.}{Secs.}
\crefname{equation}{Eq.}{Eqs.}
\renewcommand{\today}{\number\day \space \ifcase \month \or January\or February\or March\or April\or May\or June\or July\or August\or September\or October\or November\or December\fi \space \number\year} 
\def\m1r{\multicolumn{1}{r}}
\begin{document}
\title{Antiferromagnetism, spin splitting, and spin-orbit interaction in MnTe}
\author{Suman \surname{Rooj}}
\author{Jayita \surname{Chakraborty}}
\email[Email: ]{Jayita.Chakraborty1@gmail.com}
\author{Nirmal \surname{Ganguli}}
\email[Email: ]{NGanguli@iiserb.ac.in}
\affiliation{Department of Physics, Indian Institute of Science Education and Research Bhopal, Bhauri, Bhopal 462066, India}
\date{\today}
\begin{abstract}
Hexagonal MnTe emerges as a critical component in designing magnetic quantum heterostructures, calling for a detailed study. After finding a suitable combination of exchange-correlation functional and corrections, our study within {\em ab initio} density functional theory uncovers an insulating state with a preferred antiferromagnetic order. We compute the exchange interaction strengths to estimate the antiferromagnetic ordering temperature via Monte Carlo calculations. Our calculations and symmetry analysis reveal a large spin splitting in the system due to the antiferromagnetic order without considering spin-orbit interaction, except in the $k_x$-$k_y$ plane. Critically examining the band dispersion and spin textures obtained from our calculations and comparing them with an insightful symmetry analysis and analytical model, we confirm a combined Rashba-Dresselhaus interaction in the $k_x$-$k_y$ plane, around the K point of the system. Finally, we find ferroelectricity in the system for a higher energy magnetic configuration. Our results and insights would help design heterostructures of MnTe for technological applications.
\end{abstract}
\maketitle

\section{\label{sec:intro}Introduction}
Besides exhibiting interesting physical properties, magnetic quantum materials and heterostructures hold immense promise for future technology. A combination of topologically nontrivial properties with magnetism, Rashba-like spin-orbit interaction, spin Hall effect, and strong electronic correlation may lead to numerous possibilities for designing and operating devices \cite{KeimerNP17, GuiACS19}. Antiferromagnetic spintronics, where antiferromagnetic materials with substantial spin-orbit interaction can be used for superfast computational processing and nonvolatile memory by manipulating the spin-textures, may be envisaged as one such possible direction \cite{BaltzRMP18}. The idea of antiferromagnetic spintronics bases itself on the realization of externally manipulable antiferromagnetic spin textures via spin-orbit torque. Rashba-like spin-orbit interaction or spin Hall effect may help realize such tunable spin-orbit torque, storing and processing information via spin textures at a THz frequency and tiny power consumption. Recently, spin splitting due to antiferromagnetic order in the absence of combined time reversal, spatial inversion symmetry, and combined spin reversal, translation symmetry has emerged as a promising research direction \cite{YuanPRB20, YuanPRM21}. However, the search for suitable materials for the purpose is still on, inspiring researchers to design new heterostructures with the necessary ingredients \cite{ChakrabortyPRB20}. In this context, MnTe has been recognized as an interesting antiferromagnetic band insulator and a component for designing suitable heterostructures \cite{KriegnerNC16, KriegnerPRB17, MoseleyPRM22, WatanabeAPL18, YinPRL19, LiaoPRB20, WangNL21, PournaghaviPRB21}. Experiments revealed room-temperature antiferromagnetism along with anisotropic magnetoresistance and spin-flop transitions in bulk MnTe \cite{KriegnerNC16, KriegnerPRB17}. Strong magnetic anisotropy has been reported in Li-doped MnTe \cite{MoseleyPRM22}. MnTe$|$InP heterostructure exhibits interfacial conduction along with ferromagnetism \cite{WatanabeAPL18}, while thin films of antiferromagnetic MnTe reportedly host planar Hall effect \cite{YinPRL19}. Several possible heterostructures combining MnTe have been envisaged, with a few already synthesized \cite{LiaoPRB20, WangNL21, PournaghaviPRB21}. Bi$_2$Te$_3|$MnTe bilayers have been proposed to realize charge-magnon conversion, useful in antiferromagnetic magnonics to improve the performance of magnon transistors and magnon torque memories \cite{LiaoPRB20}. Ferromagnetic$|$antiferromagnetic$|$ferromagnetic heterostructures of Cr$_2$Te$_3|$MnTe$|$Cr$_2$Te$_3$ has been synthesized to host high coercivity and exchange bias \cite{WangNL21}. \citet{PournaghaviPRB21} proposed Chern insulator and axion insulator phases in MnTe$|$Bi$_2$(Se,Te)$_3|$MnTe heterostructures based on density functional theory (DFT) and tight-binding model calculations. A similar compound, GeTe, has been reported from DFT calculations to host a strong Rashba interaction along with ferroelectricity in the bulk form, where the Te atoms play a major role in realizing the spin-orbit interaction \cite{DiSanteAM13}. Based on DFT calculations, bulk MnTe has recently been shown to host a spin splitting even without considering spin-orbit interaction, a feature that may offer more robust application potential than the spin splitting due to spin-orbit interaction \cite{SmejkalPRX22}.

The above discussion reveals the usefulness of MnTe for quantum technologies. It highlights the importance of thoroughly understanding the physical properties, particularly the electronic structure, magnetic properties, and spin-orbit interaction, so that heterostructures involving MnTe can be rationally designed to host the desired properties. Therefore, using first principles density functional theory within a judicially chosen combination of exchange-correlation functional and corrections, we study the interesting physical properties of MnTe in its hexagonal bulk structure. After carefully understanding the electronic structure, preferred magnetic configuration, and magnetic interactions, we critically examine the spin splitting without and with spin-orbit interaction using DFT calculations, symmetry analysis, and an analytical model. Subsequently, we explore the possibility of ferroelectricity in the system. The remainder of the article is organized as follows: The crystal structure and our calculation methodologies are described in \cref{sec:method}. The results of our calculations are thoroughly discussed in \cref{sec:results}. Finally, we summarize the work in \cref{sec:conc}.

\section{\label{sec:method}Crystal structure and Method}
Bulk MnTe crystallizes in a hexagonal NiAs structure with space group $P6_{3}/mmc$, Mn and Te occupying $2a$ and $2c$ Wyckoff positions, respectively. \Cref{fig:MagConfig}(a) depicts a unit cell of MnTe, illustrating face-sharing MnTe$_6$ octahedra. The total energy, electronic structure, magnetic properties, spin-orbit interaction, and ferroelectricity calculations presented here are performed within density functional theory, as implemented in the {\scshape vasp} code \cite{vasp1, vasp2}. A plane wave basis set with 500~eV kinetic energy cutoff is employed to expand the wavefunctions within the projector augmented wave (PAW) method \cite{paw}. We systematically tested various exchange-correlation functionals combined with appropriate corrections including local density approximation (LDA) \cite{ldaCA, PerdewPRB81}, generalized gradient approximation (GGA) \cite{pbe}, a meta-GGA functional {\em viz.} SCAN \cite{scan}, Hubbard-$U$ correction \cite{DudarevPRB98}, and van der Waals correction rVV10 \cite{rVV10} to find which combination of functional and corrections best suite our purpose. The Brillouin zone integration is performed within corrected tetrahedron method \cite{BlochlPRB94T} using a $\Gamma$-centered $k$-point mesh of $9 \times 9 \times 6$ and $5 \times 5 \times 4$ for the unit cell and a $2 \times 2 \times 2$ supercell, respectively. A $U_\text{eff} = U - J =3$~eV is used for the description of Mn-$3d$ states with reasonably strong Coulomb correlation. The atomic positions and lattice vectors are optimized to minimize the Hellman-Feynman force on each atom to a threshold of $10^{-2}$~eV~\AA$^{-1}$ and the stress on the simulation cell, respectively.
\begin{table}
\caption{\label{tab:LatticeConstant}Lattice constants of MnTe obtained from the experiment and calculations using different exchange-correlation functionals and corrections are tabulated here.}
\begin{ruledtabular}
    \begin{tabular}{l..}
     Experiment and different functionals & \multicolumn{1}{r}{$a = b$~(\AA)} & \multicolumn{1}{r}{$c$~(\AA)} \\
    \hline
    Experiment \cite{Szuszkiewicz2005} & 4.15 & 6.71 \\
    Spin-unpolarized LDA & 4.002 & 4.802 \\
    Spin-unpolarized LDA + $U$ & 4.021 & 4.849 \\
    Spin-polarized LDA & 3.795 & 5.828 \\
    Spin-polarized LDA + $U$ & 4.067 & 6.436 \\
    Spin-unpolarized GGA & 4.097 & 4.899 \\
    Spin-unpolarized GGA + $U$ & 4.128 & 5.020 \\
    Spin-polarized GGA & 4.102 & 6.444 \\
    Spin-polarized GGA + $U$ & 4.193 & 6.704\\
    Spin-unpolarized SCAN + rVV10 & 4.041 & 4.861 \\
    Spin-unpolarized SCAN + rVV10 + $U$ & 4.101 & 5.105\\
    Spin-polarized SCAN + rVV10 & 4.079 & 6.493 \\
    Spin-polarized SCAN + rVV10 + $U$ & 4.165 & 6.683 \\
    \end{tabular}
\end{ruledtabular}
\end{table}
\cref{tab:LatticeConstant} compares the predictive capability of different exchange-correlation functionals and corrections, revealing spin-polarized SCAN + rVV10 + Hubbard-$U$ and spin-polarized GGA + Hubbard-$U$ functionals to best predict the lattice constants. Since a combination of SCAN and rVV10 has been demonstrated to outperform other functionals in layered materials \cite{rVV10}, we continue our further calculations within the framework of SCAN + rVV10 + Hubbard-$U$. Whenever appropriate, spin-orbit interaction is considered in our calculations. The Berry-phase method is employed to calculate the ferroelectric properties of the system \cite{Resta1992, Vanderbilt1993}.

\section{\label{sec:results}Results and discussions}
After choosing an appropriate combination of exchange-correlation functional and corrections, we systematically investigate the electronic structure, magnetic properties, spin-orbit interaction, and ferroelectric properties of MnTe.
\subsection{Electronic Structure and magnetic properties}
\begin{figure}
  \includegraphics[scale=0.4]{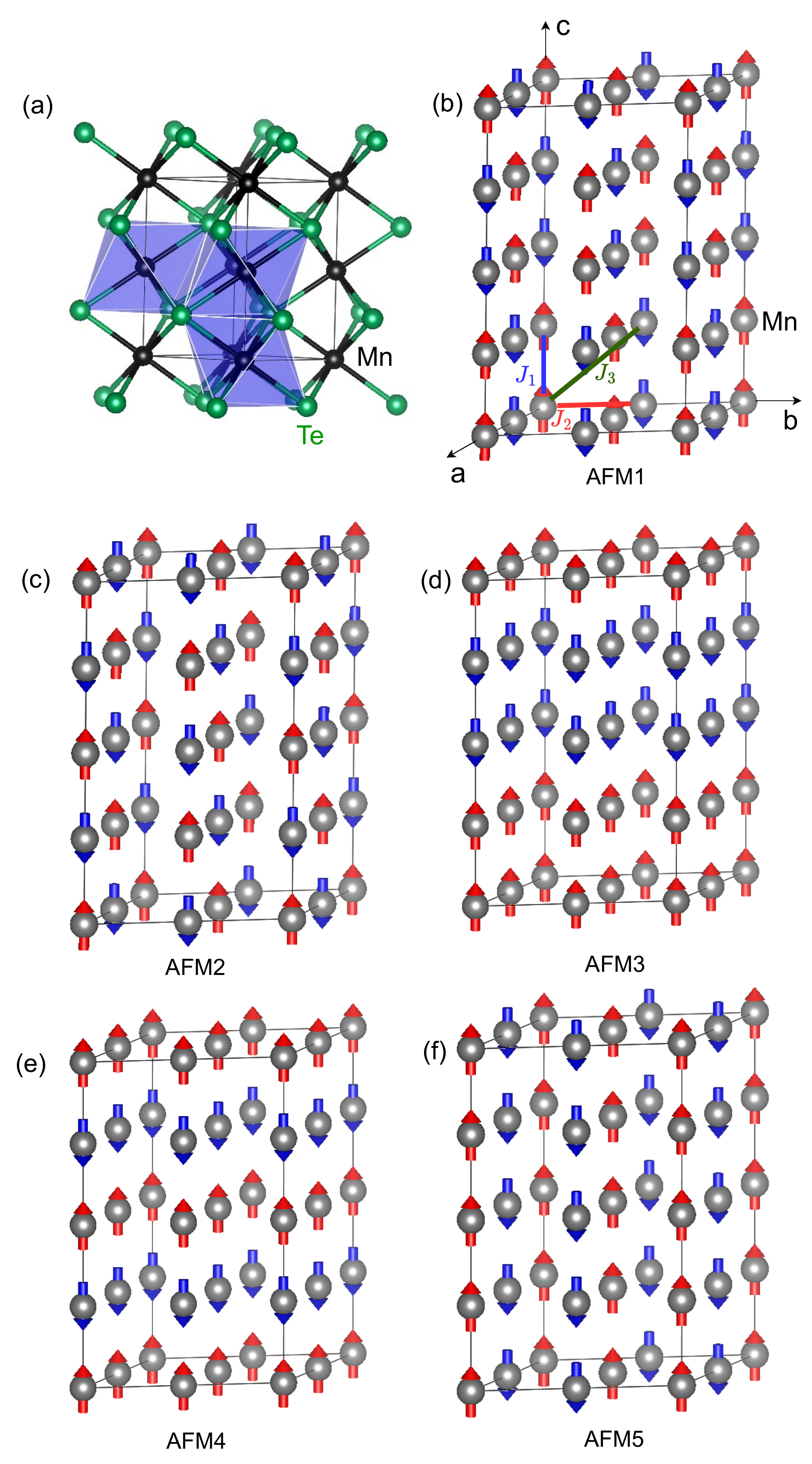}
 \caption{\label{fig:MagConfig}The face sharing MnTe$_6$ octahedra are illustrated in the depiction of a unit cell in (a), while The magnetic exchange paths and the possible antiferromagnetic configurations AFM1, AFM2, AFM3, AFM4, and AFM5 are depicted in (b), (c), (d), (e), and (f), respectively. The illustrations have been prepared using {\scshape vesta} software \cite{VESTA}.}
\end{figure}
\begin{figure}
    \includegraphics[scale=0.36]{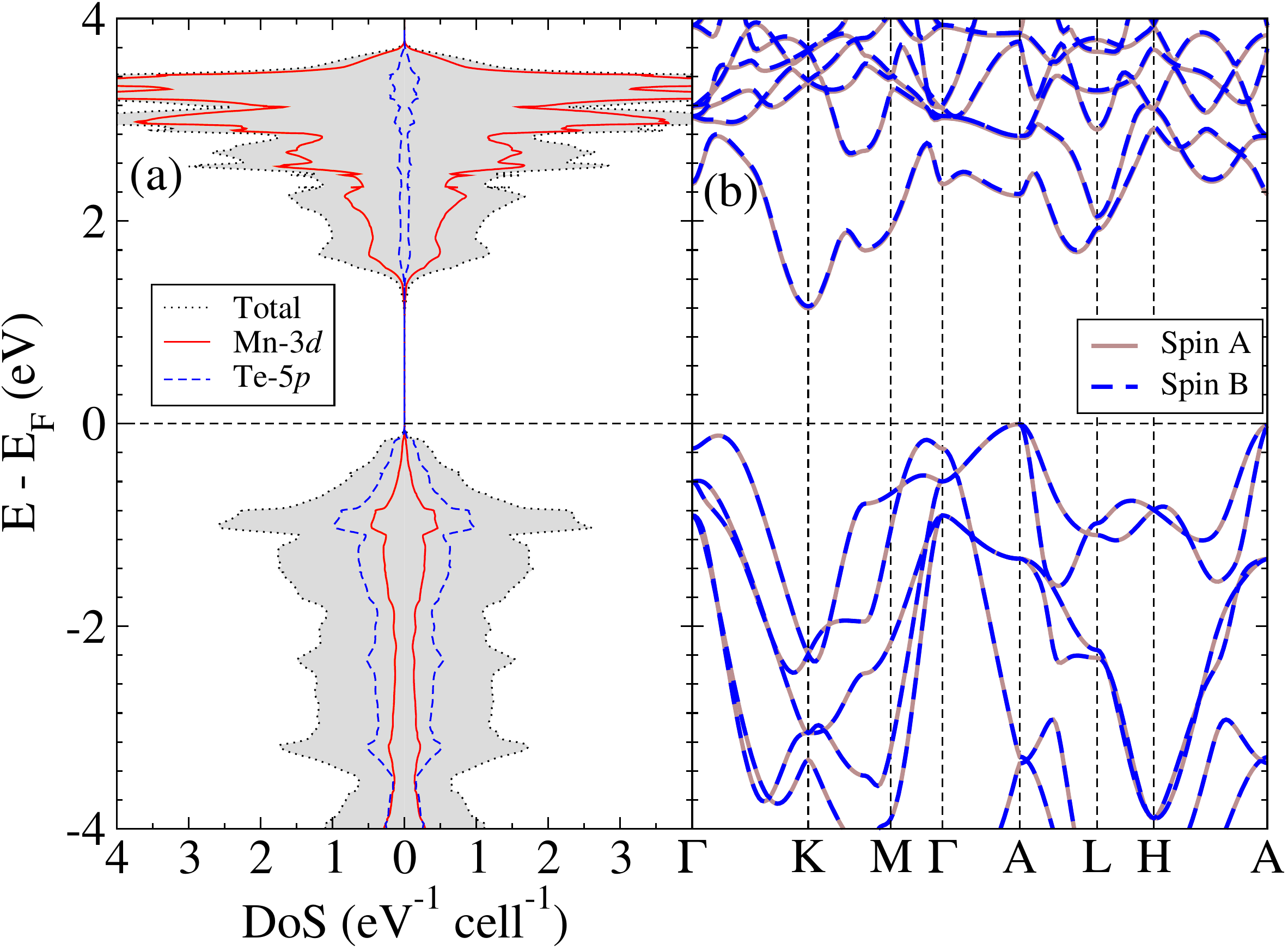}
    \caption{\label{fig:DoSspin}Spin-polarized density of states along with Mn-$3d$ and Te-$5p$ orbital-projected DoS for AFM4 configuration is displayed in (a), while (b) represents the band dispersion from both spins marked as spin A and spin B.}
\end{figure}

In a nominal sense of ionic compound, Mn and Te atoms in MnTe are expected in $2+$ and $2-$ oxidation states, respectively. We start with defining the possible magnetic configurations and identifying the exchange paths in the system. Besides the ferromagnetic (FM) configuration, where magnetic moments from all Mn atoms align along the same direction, several antiferromagnetic (AFM) configurations may be envisaged depending on the exchange interactions, leading to no net magnetic moment of the system. The arrangement of magnetic moments in AFM1, AFM2, AFM3, AFM4, and AFM5 are illustrated in \cref{fig:MagConfig}(b),(c),(d),(e),(f), respectively. \cref{tab:RelativeEnergy} shows the relative energies for all the magnetic configurations, suggesting AFM4 to have the lowest energy. The magnetic moments of Mn ions in AFM4 configuration are parallel to each other in the hexagonal basal plane, whereas antiparallel along the $c$-direction. Unless stated otherwise, our results discussed subsequently corresponds to the lowest energy AFM4 configuration.
\begin{table}
\caption{\label{tab:RelativeEnergy}The relative energies for different magnetic configurations are tabulated here.}
\begin{ruledtabular}
    \begin{tabular}{l.}
    Magnetic configuration & \multicolumn{1}{c}{Relative energy (meV)} \\
    \hline
    FM & 938.87 \\
    AFM1 & 379.46 \\
    AFM2 & 26.69 \\
    AFM3 & 400.44 \\
    AFM4 & 0.0 \\
    AFM5 & 769.39 \\
    \end{tabular}
\end{ruledtabular}
\end{table}
The density of states (DoS) for both spins and band dispersion for one spin of the system in AFM4 configuration, shown in \cref{fig:DoSspin}(a) and \cref{fig:DoSspin}(b), respectively, suggest a band gap of $\sim$1.16~eV, which is in reasonable agreement with 1.27~eV gap reported from experiment \cite{SzuszkiewiczPRB06}. We attribute our choice of SCAN + rVV10 + Hubbard-$U$ framework to this fair agreement. The projected DoS (see \cref{fig:DoSspin}(a)) indicates that while the valence band comprises hybridized Mn-$3d$ and Te-$5p$ orbitals, the conduction band predominantly consists of only Mn-$3d$ orbitals.
The band dispersion (see \cref{fig:DoSspin}(b)) reveals an indirect band gap with the valence band maximum at A-point and the conduction band minimum at K-point. A projected magnetic moment of 4.54~$\mu_B$ is observed at each Mn, suggesting a high-spin configuration of Mn$^{2+}$ ion. We note that the projected magnetic moment is often slightly less than the actual magnetic moment when a plane wave basis set is used for expanding the wavefunctions. Neutron diffraction experiments revealed a magnetic moment of 4.7$-$5~$\mu_B$ per Mn atom \cite{KriegnerPRB17}. The band dispersion shows a perfect overlap of the bands of both spins along the chosen high-symmetry directions.

\subsubsection{Exchange interaction and ordering temperature}
The magnetic exchange interaction strengths $J$ in the spin Hamiltonian $H = - \sum_{ij} J_{ij}~S_i S_j$ with $S$ and $i,j$ representing spins and site indices, respectively, can be evaluated along different directions from the relative energies of the various magnetic configurations listed in \cref{tab:RelativeEnergy} using a method vividly described in ref.~\cite{ChakrabortyJPCM17, ChakrabortyPRB18}.
\begin{table}
\caption{\label{tab:ExcInteraction}The exchange interaction parameters ($J$s) are tabulated here.}
\begin{ruledtabular}
    \begin{tabular}{l..}
    Exchange path &  \multicolumn{1}{c}{Distance (\AA)} &  \multicolumn{1}{c}{Exchange strength (meV)} \\
    \hline
    $ J_1 $ & 3.342 & -3.79 \\
    $ J_2 $ & 4.165 & -0.02 \\
    $ J_3 $ & 5.34 & -0.04\\ 
    \end{tabular}
\end{ruledtabular}
\end{table}
The important exchange interaction strengths along the paths marked as $J_1$, $J_2$, and $J_3$ marked in \cref{fig:MagConfig}(b) are estimated and tabulated in \cref{tab:ExcInteraction}. While the Mn ions connected via $J_1$ path can have direct exchange interaction, $J_2$ and $J_3$ paths allow superexchange via one and two Te-ions, respectively, explaining the different orders of their magnitude. The negative signs of the prominent $J$-values suggest an antiferromagnetic ground state, consistent with our findings. 

A magnetic ordering temperature, known as the N\'{e}el temperature for antiferromagnetic systems, may be estimated using the exchange interaction strengths given in \cref{tab:ExcInteraction}. We compute the magnetic specific heat per spin $C$ given as \cite{AlbaalbakyPRB17}
\begin{equation}
    C = \frac{1}{N} \frac{\partial U}{\partial T} = \frac{\langle \varepsilon^2 \rangle - \langle \varepsilon \rangle^2}{N k_B T^2}, \label{eq:specific}
\end{equation}
where $\varepsilon$, $N$, $k_B$, and $T$ represent the energy of each magnetic configuration, number of spins, Boltzmann constant, and temperature, respectively. $U(T) = \langle \varepsilon \rangle$ is the average internal energy, $\langle ... \rangle$ representing thermal average at a given temperature. We calculate the energy of magnetic configurations as
\begin{equation}
    \varepsilon = -J_1 \sum_{ij} {S_i}{S_j} - J_3 \sum_{ij} {S_i}{S_j}, \label{eq:spinEnergy}
\end{equation}
where $i$, $j$ run over the atoms satisfying the distance of the corresponding exchange path $J_1$, or $J_3$. Owing to its small value, we ignore $J_2$ for our calculations but retain $J_3$ due to twice as many neighbors at that distance compared to $J_2$, making it significant. A large projected magnetic moment at the Mn site indicates the high spin configuration of Mn$^{2+}$ ions, justifying our approach. Considering three different lattice sizes $15 \times 15 \times 15$, $18 \times 18 \times 18$, and $20 \times 20 \times 20$ within periodic boundary condition, we evolved the system in imaginary time using Metropolis Monte Carlo simulation algorithm and Boltzmann distribution function to calculate $\langle \varepsilon^2 \rangle$ and $\langle \varepsilon \rangle$ at different temperatures. A spin $S_i$ is randomly selected from the lattice, and the change in energy $\Delta \varepsilon$ upon flipping the spin is computed from \cref{eq:spinEnergy}. If the corresponding Boltzmann weight $\exp(-\Delta \varepsilon/k_BT)$ is greater than a uniform random number $r \in (0,1)$, the spin-flip is accepted. Subsequently, we move on to the next time step by randomly choosing another spin $S_i$ \cite{GanguliPRBR19, Ganguli2PRB19}. The averaging is performed over $10^5$ imaginary time steps after bringing the system to thermal equilibrium in $10^7$ imaginary time steps for a given temperature.
\begin{figure}
    \includegraphics[width=8.5cm, height=6cm]{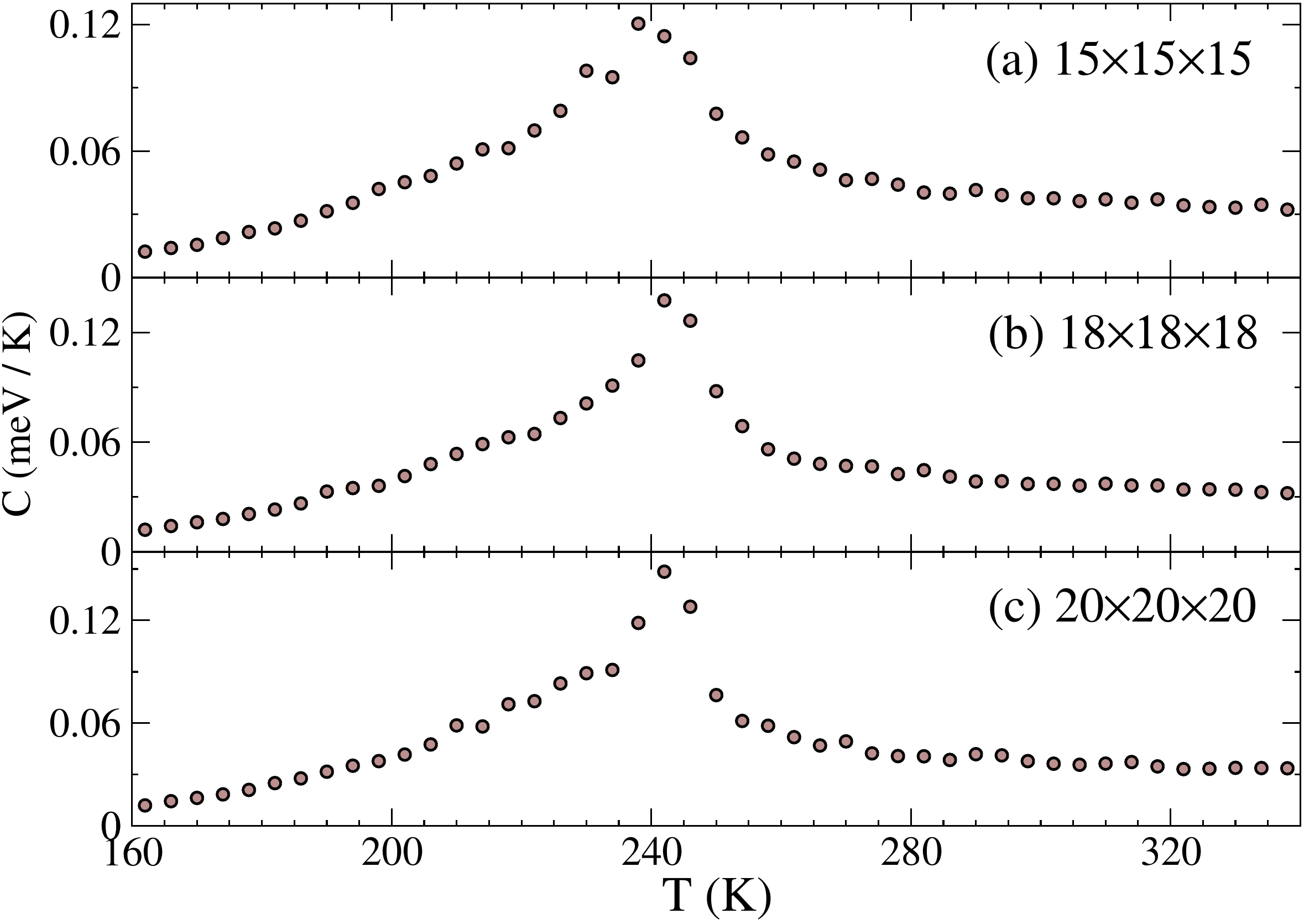}
    \caption{\label{fig:specificHeat}The magnetic specific heat with temperature for lattice sizes (a) $15 \times 15 \times 15$, (b) $18 \times 18 \times 18$, and (c) $20 \times 20 \times 20$ are shown here.}
\end{figure}
Our results, displayed in \cref{fig:specificHeat} for lattice sizes $15 \times 15 \times 15$, $18 \times 18 \times 18$, and $20 \times 20 \times 20$ lattice sizes within periodic boundary condition, indicate an ordering temperature of $\sim$240~K with negligible variation for different lattice sizes. We note that experiments reported a magnetic ordering temperature around 310~K \cite{KriegnerPRB17}, indicating a minor underestimate from our calculations, possibly due to errors in the calculated $J$-values.

\subsubsection{Spin splitting}
\begin{figure*}
	\includegraphics[scale = 0.64]{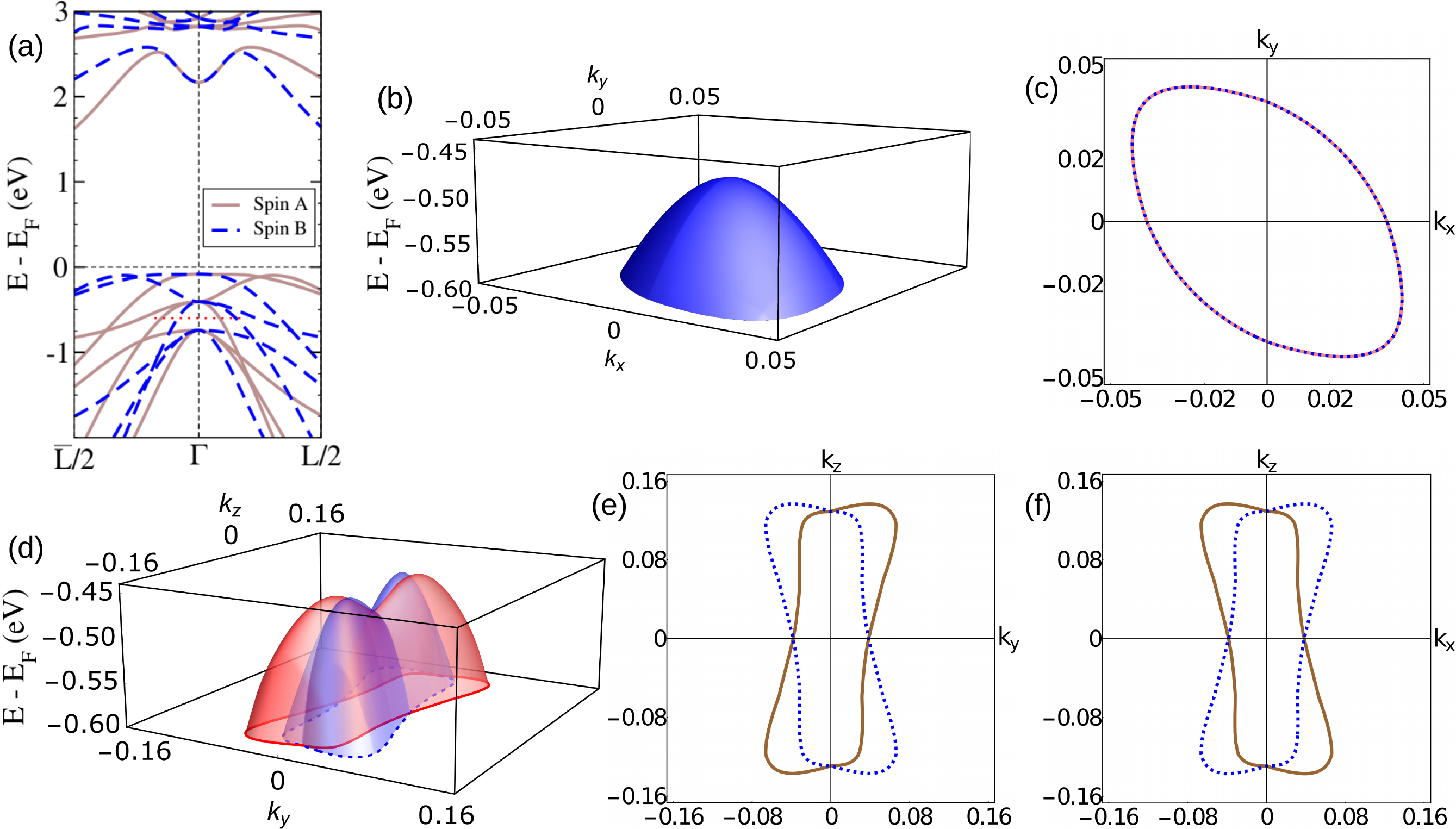}
	\caption{\label{fig:AFMspinSplitting}Panel (a) shows the spin-polarized MnTe bands along $\bar{L}/2 \to \Gamma \to L/2$ direction. We identify a pair of such spin-split bands within a small energy range $[-0.6,-0.44]$~eV relative to the Fermi level, intersecting a horizontal dotted (red) line at $E - E_F = -0.6$~eV, and show their dispersion as functions of $(k_x, k_y)$ with $k_z = 0$ in (b). The corresponding isoenergetic contours in the $k_z = 0$ plane are shown in (c) for $E - E_F = -0.6$~eV. Panel (d) shows the dispersion of the same pair of bands as functions of $(k_y, k_z)$ with $k_x = 0$. Panels (e) and (f) exhibit the isoenergetic contours for $E - E_F = -0.6$~eV in the $k_x = 0$ and $k_y = 0$ planes, respectively.}
\end{figure*}
After confirming the magnetic ground state and estimating the magnetic exchange interactions, hence the ordering temperature, we analyze the spin splitting in the system driven by antiferromagnetic exchange interaction \cite{SmejkalPRX22}. While the band dispersion shown in \cref{fig:DoSspin}(b) along $\Gamma \to K \to M \to \Gamma \to A \to L \to H \to A$ direction reveals no spin splitting, the band dispersion plotted along $\bar{L}/2 \to \Gamma \to L/2$ direction, displayed in \cref{fig:AFMspinSplitting}(a) without considering spin-orbit interaction reveals pronounced spin splitting, in agreement with Ref.~\cite{SmejkalPRX22}, calling for a careful analysis of the spin splitting. The magnetic system in AFM4 configuration does not preserve the combined time-reversal and spatial inversion symmetry; a combined spin rotation (reversal) and translation symmetry is also broken, except for some local symmetries in the $k_x$-$k_y$ plane, owing to the parallel arrangement of spins in the $ab$ plane and antiparallel arrangement along the $c$-direction. Thus, the spin degeneracy remains protected in the $k_x$-$k_y$ plane \cite{YuanPRB20}, as seen from \cref{fig:AFMspinSplitting}(b) and \cref{fig:AFMspinSplitting}(c), exhibiting the 3D band dispersion as a function of $(k_x, k_y)$ and the isoenergetic contours for $E - E_F = -0.6$~eV in the $k_x$-$k_y$ plane, respectively, for the pair of bands within a small energy range $[-0.6,-0.44]$~eV relative to the Fermi level, intersecting a horizontal dotted (red) line at $E - E_F = -0.6$~eV in \cref{fig:AFMspinSplitting}(a). However, since the local symmetry does not hold in other planes, the 3D band dispersion as a function of $(k_y, k_z)$ and the corresponding isoenergetic contours for $E - E_F = -0.6$~eV in the $k_y$-$k_z$ plane, shown in \cref{fig:AFMspinSplitting}(d) and \cref{fig:AFMspinSplitting}(e), respectively, for the same pair of bands exhibit a pronounced spin splitting without considering spin-orbit interaction. Similarly, the isoenergetic contours in the $k_x$-$k_z$ plane displayed in \cref{fig:AFMspinSplitting}(f) also exhibit large spin splitting, owing to the absence of the local symmetry in the plane. We note that the spin splitting without spin-orbit interaction discussed here may not be useful for spintronic applications in a heterostructure involving (0001)-terminated MnTe unless the magnetic configuration drastically changes upon forming the heterostructure since the Brillouin zone would effectively collapse to a $k_x$-$k_y$ plane in such a heterostructure. However, if realized, a spin splitting in the $k_x$-$k_y$ plane due to spin-orbit interaction may be useful for spintronic applications in a heterostructure involving (0001)-terminated MnTe, encouraging us to investigate the implications of spin-orbit interaction in the system under consideration.

\subsection{Spin-orbit interaction}
Upon considering spin-orbit interaction in our DFT calculations, we find substantial magneto-crystalline anisotropy and splitting of bands in the momentum space due to Rashba-like interaction along with a projected orbital moment of 0.011~$\mu_B$ at each Mn site. Below we discuss our results in detail.

\subsubsection{Magneto-crystalline anisotropy}
Spin-orbit interaction often leads to anisotropy in spin quantization along different crystallographic directions, particularly for non-cubic crystals. We compute the relative energies for quantizing the spin along four crystallographic directions {\em viz.} (1000), (0100), (0001), and $(11\bar{2}0)$ for hexagonal MnTe; and tabulate the results in \cref{tab:anisotropy}.
\begin{table}
\caption{\label{tab:anisotropy}The relative energies of spin quantization along different crystallographic directions considering spin-orbit interaction are tabulated here.}
\begin{ruledtabular}
    \begin{tabular}{l.}
    Spin quantization axis & \multicolumn{1}{c}{Energy~(meV/formula unit)} \\
    \hline
    $\left(1000\right)$ & 0.19 \\
    $\left(0100\right)$ & 3.23 \\ 
    $\left(0001\right)$ & 0.075 \\
    $\left(11\Bar{2}0\right)$ & 0 \\
    \end{tabular}
\end{ruledtabular}
\end{table}
Our results reveal $(11\bar{2}0)$ direction in $ab$-plane to be the most preferred spin quantization direction among the ones considered, consistent with a previous report of magneto-crystalline anisotropy in hexagonal MnTe \cite{KriegnerPRB17}, resulting in a collinear magnetic arrangement (AFM4 configuration), leaving the magnetic space group unchanged.

\subsubsection{Rashba-like interaction}
\begin{figure}
 \includegraphics[scale=0.36]{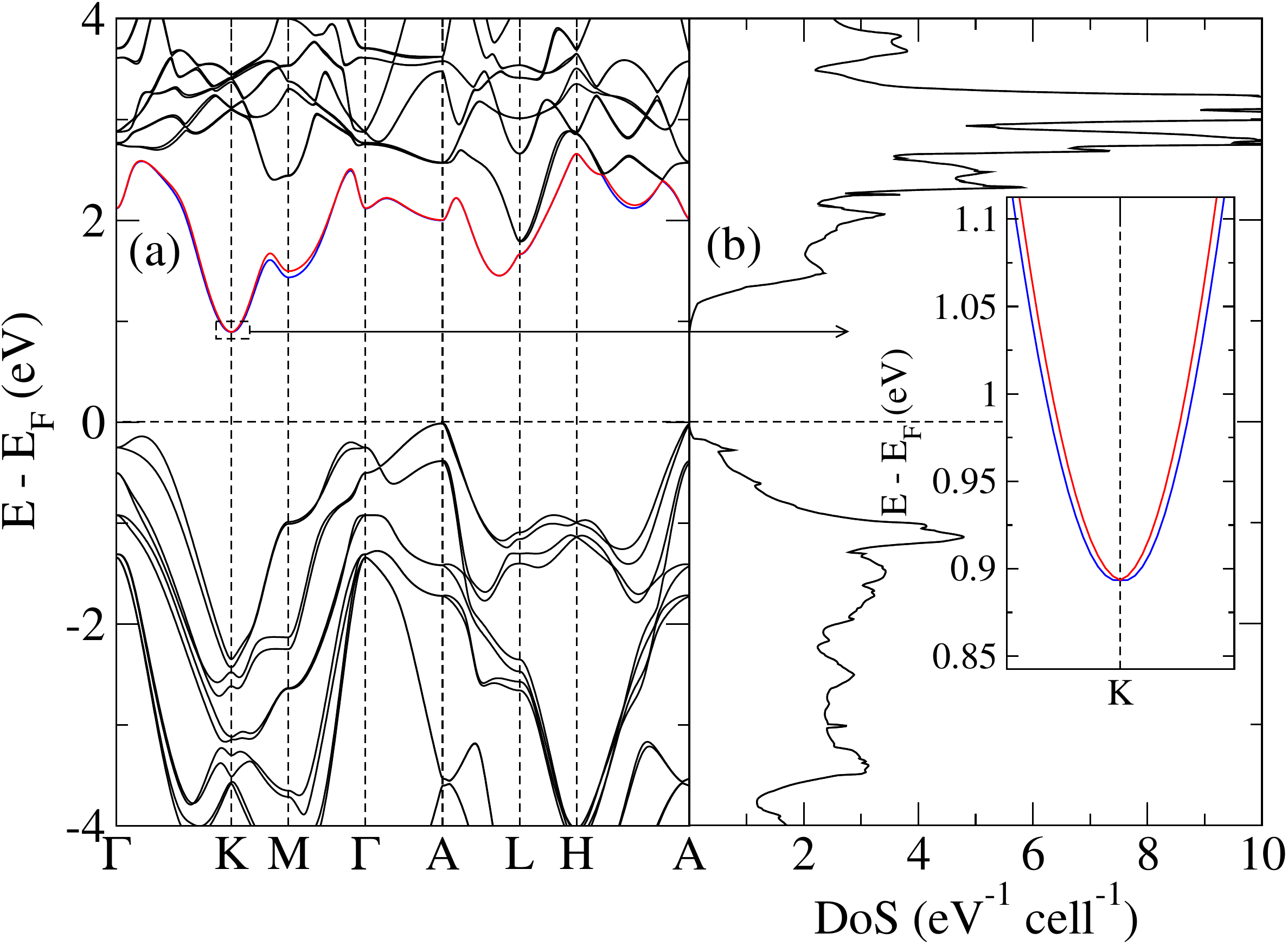}
 \caption{\label{fig:soiBandDoS}The band structure and density of states of MnTe considering spin-orbit interaction are shown in (a) and (b), respectively. The inset of (b) enlarges the bottom of the conduction band at the high-symmetry point K to reveal Rashba-like splittings of the bands in momentum space.}
\end{figure}
In order to understand the electronic structure of MnTe upon considering spin-orbit interaction, we plot the band dispersion along the high-symmetry lines of a hexagonal Brillouin zone and density of states, as shown in \cref{fig:soiBandDoS}(a) and \cref{fig:soiBandDoS}(b), respectively. We note that considering spin-orbit interaction has reduced the calculated band gap to $\sim$0.9~eV. Some bands below and above the Fermi level exhibit a Rashba-like spin splitting in the momentum space; we take a closer look at the bottom of the conduction band having a predominant Mn-$3d$ character, as featured in the inset of \cref{fig:soiBandDoS}(b).

\begin{figure*}
 \includegraphics[width=17.88 cm, height=9.7 cm]{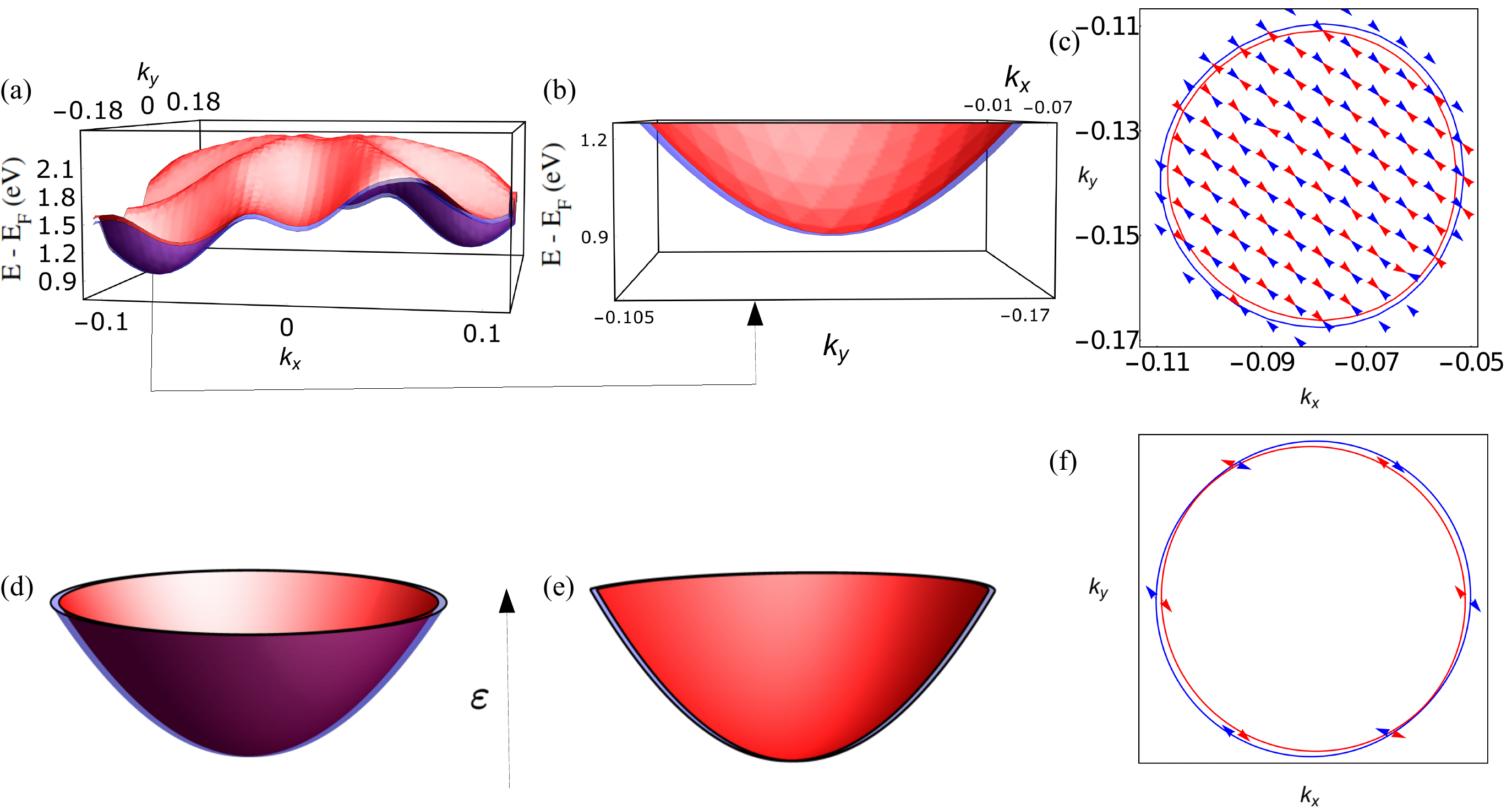}
 \caption{\label{fig:3DModel}The energy dispersion as functions of $(k_x, k_y)$ (3D bands) for the lowest two conduction bands is shown in panel (a). Panel (b) exhibits a cross-section of the 3D bands near the high-symmetry point K. Isoenergetic contours for the same bands at $E - E_F = 1.20$~eV along with projected spins over an energy range $[0.88, 1.25]$~eV relative to the Fermi level from DFT calculations are displayed in (c). Panels (d), (e), and (f) represent the 3D bands, a cross-section of the 3D bands, and isoenergetic contours with projected spins, respectively, obtained from our combined Rashba-Dresselhaus interaction model.}
\end{figure*}
To better understand the spin-orbit interaction in the lowest pair of conduction bands, we plot the energy dispersion as functions of $(k_x, k_y)$, as displayed in \cref{fig:3DModel}(a). A cross-section of the bands in an appropriate part of the Brillouin zone, as displayed in \cref{fig:3DModel}(b), reveals Rashba-like splitting of the pair of bands in momentum space. \cref{fig:3DModel}(c) exhibits a pair of isoenergetic contours for the same pair of bands at $E - E_F = 1.20$~eV along with projected spin texture, as obtained over an energy range $[0.88, 1.25]$ relative to the Fermi level from our DFT calculations \cite{KumarPRB22}. We find the projected spins arranging themselves like a persistent spin helix (PSH) \cite{KoralekN09}, a characteristic of combined Rashba and Dresselhaus spin-orbit interactions of similar strengths.

The idea of Rashba-Dresselhaus interaction in bulk MnTe intrigues us as the structure corresponds to a centrosymmetric space group $P6_3/mmc$. Dresselhaus and Rashba interactions require bulk inversion asymmetry (BIA) and structure inversion asymmetry (SIA) with a microscopic electric field, respectively. The asymmetry may arise from the point groups corresponding to some atomic sites instead of the space group \cite{ZhangNP14, FajardoPRB19, AcostaPRB21, WinklerSOC03}. In the present case, Mn and Te ions occupy $2a$ and $2c$ Wyckoff positions corresponding to point groups $D_{3d}$ and $D_{3h}$, respectively. While the point group $D_{3d}$ corresponding to Mn-sites preserves centrosymmetry, the point group $D_{3h}$ corresponding to Te-sites, although nonpolar, lacks centrosymmetry. The bulk inversion asymmetry thus introduced is expected to lead to only a Dresselhaus (D-2) interaction \cite{ZhangNP14}. Additionally, the microscopic arrangement of alternating $+2|-2$ charged layers along the $(0001)$ direction due to the arrangement of Mn$^{2+}$ and Te$^{2-}$ ions leads to a microscopic electric field along the same direction \cite{GanguliPRL14}, resulting in Rashba interaction. Hence, finding a combined Rashba-Dresselhaus spin-orbit interaction in the system is possible. Further, the crystal point group symmetry coincides with the wave vector point group symmetry $D_{3h}$ at the high-symmetry K point \cite{DresselhausGroupTheory08}. We identify a magnetic space group $Cmcm$ for our system \cite{GallegoJAC16}, a type~I magnetic space group according to Ref.~\cite{BradleySymmetrySolid} and spin splitting prototype 4A according to Ref.~\cite{YuanPRM21}, where spin splitting is expected even without spin-orbit interaction. However, owing to a collinear magnetic arrangement, the local symmetry discussed earlier for the $k_x$-$k_y$ plane ensures that the spin expectation value $\vec{S}_n(\vec{k})$ for the $n^\text{th}$ band satisfies the condition $\vec{S}_n(-\vec{k}) = -\vec{S}_n(\vec{k})$ in the $k_x$-$k_y$ plane around the K point, as verified from the spin texture displayed in \cref{fig:3DModel}(c). Hence, the Rashba-Dresselhaus terms would take their conventional form in the Hamiltonian for the $k_x$-$k_y$ plane around the K point \cite{AcostaPRB21}.

To examine the hypothesis of Rashba-Dresselhaus interactions against our DFT results, considering the above-mentioned symmetry conditions, we construct an analytical model of combined Rashba-Dresselhaus spin-orbit interaction in the $k_x$-$k_y$ plane around the K point, as described by the Hamiltonian $H_\text{RD}$ (in the units of $\hbar$) \cite{ChakrabortyPRB20}
\begin{align}
    H_\text{RD} &= H_0 + H_\text{R} + H_\text{D}, ~~\text{where} \label{eq:ardH} \\
    H_0 &= -\frac{1}{2m^*} \left( \frac{\partial^2}{\partial x^2} + \frac{\partial^2}{\partial y^2} \right) ~~\text{and} \nonumber \\
    H_\text{R} &= \alpha (k_y \sigma_x - k_x \sigma_y); ~~ H_\text{D} = \beta (k_y \sigma_y - k_x \sigma_x), \nonumber
\end{align}
with $(\sigma_x, \sigma_y)$ and $m^*$ representing the Pauli matrices and the effective mass, respectively. Considering the Rashba and Dresselhaus terms as perturbations to the free electron-like Hamiltonian $H_0$, we find the energy eigenvalues
\begin{align}
    \varepsilon_\text{RD}^{\pm}(k_x, k_y) =& \frac{k_x^2 + k_y^2}{2m^*} \pm \sqrt{(\alpha^2 + \beta^2)(k_x^2 + k_y^2) - 4 \alpha \beta k_x k_y} \nonumber \\
    =& \frac{k_{\parallel}^2}{2m^*} \pm k_{\parallel} \sqrt{\alpha^2 + \beta^2 - 2 \alpha \beta \sin 2\phi}, \label{eq:eigenvalue}
\end{align}
where we denote $(k_{\parallel}, \phi)$ as the polar coordinates in the $k_x$-$k_y$ plane. Setting the expressions for eigenvalues in \cref{eq:eigenvalue} to a suitable constant energy will give us the equations for isoenergetic contours in $k_x$-$k_y$ plane. The eigenstates for the Hamiltonian may be expressed as
\begin{align}
	| \pm \rangle_\text{RD} &= \frac{1}{\sqrt{2}} \left( \pm \zeta_\text{RD} |\uparrow \rangle + |\downarrow \rangle \right), ~~ \text{where} \label{eq:eigenstates} \\
	\zeta_\text{RD} &= i \frac{\sqrt{\alpha^2 + \beta^2 - 2 \alpha \beta \sin 2\phi}}{\alpha \exp(i \phi) - i \beta \exp(-i \phi)}, ~ \zeta_\text{RD}^* \zeta_\text{RD} = 1. \nonumber
\end{align}
The projected spin components $\langle S_x \rangle$ and $\langle S_y \rangle$ may be evaluated using the eigenstates in \cref{eq:eigenstates} as
\begin{align}
	\langle S_x \rangle_\text{RD}^+ &= \frac{1}{2}~ _\text{RD}\langle + | \sigma_x | + \rangle_\text{RD} = \frac{1}{4} (\zeta_\text{RD} + \zeta_\text{RD}^*) \nonumber \\
	&= \frac{1}{2} \frac{\alpha \sin \phi - \beta \cos \phi}{\sqrt{\alpha^2 + \beta^2 - 2 \alpha \beta \sin 2\phi}} \nonumber \\
	&= -\frac{1}{2}~ _\text{RD}\langle - | \sigma_x | - \rangle_\text{RD} = -\langle S_x \rangle_\text{RD}^-, \nonumber \\
	\langle S_y \rangle_\text{RD}^+ &= \frac{1}{2}~ _\text{RD}\langle + | \sigma_y | + \rangle_\text{RD} = \frac{1}{4} (\zeta_\text{RD} - \zeta_\text{RD}^*) \nonumber \\
	&= \frac{1}{2} \frac{\beta \sin \phi - \alpha \cos \phi}{\sqrt{\alpha^2 + \beta^2 - 2 \alpha \beta \sin 2\phi}} \nonumber \\
	&= -\frac{1}{2}~ _\text{RD}\langle - | \sigma_y | - \rangle_\text{RD} = -\langle S_y \rangle_\text{RD}^- \nonumber \\
	\langle S_z \rangle_\text{RD}^+ &= \frac{1}{2}~ _\text{RD}\langle + | \sigma_z | + \rangle_\text{RD} = \frac{1}{4} (\zeta_\text{RD}^* \zeta_\text{RD} -1) \nonumber \\
	&= 0 = \frac{1}{2}~ _\text{RD}\langle - | \sigma_z | - \rangle_\text{RD} = \langle S_z \rangle_\text{RD}^-. \label{eq:RDspin}
\end{align}
\cref{fig:3DModel}(d) and \cref{fig:3DModel}(e) illustrate the eigenvalues obtained in \cref{eq:eigenvalue}, while \cref{fig:3DModel}(f) depict the corresponding isoenergetic contours for a suitable energy and the projected spin directions obtained from \cref{eq:RDspin}. We note the remarkable match between our DFT results and a similar depiction from the analytical model considered here, suggesting that our model essentially describes the spin-orbit interaction observed in the system. The model reasonably fits the DFT results for $m^* = 0.003 m_e$, $m_e$ denoting the mass of an electron, $\alpha = 0.39$~eV~\AA, and $\beta = -0.30$~eV~\AA, thus conclusively characterizing the nature of spin-orbit interaction in the system. The remarkably small value of the effective mass $m^*$ indicates a super-light electron-like band under consideration.

\subsection{Magnetostriction and ferroelectricity}
Finally, we explore the possibility of ferroelectric polarization in hexagonal bulk MnTe system with and without spin-orbit interaction. Our results reveal that upon optimizing, the atomic positions of the lowest energy AFM4 magnetic configuration do not change from the Wyckoff $2a$ and $2c$ positions. As discussed earlier, the $P6_3/mmc$ space group preserved overall centrosymmetry and none of the $D_{3d}$ and $D_{3h}$ point groups associated with Mn and Te sites, respectively, exhibit any polar nature. Hence, as expected, our calculations reveal no ferroelectric polarization for bulk MnTe in the AFM4 configuration. However, we observe a pronounced magnetostrictive effect for AFM1 configuration with the Mn magnetic moments arranged in $\uparrow \uparrow \downarrow \downarrow$ order along the $c$ direction (see \cref{fig:MagConfig}(b)) \cite{ChakrabortyJMMM19}. Upon relaxation, the distance between Mn atoms along the $c$-direction with parallel (antiparallel) spin orientation increases (decreases) by $\sim$0.025~\AA. Such structural changes due to magnetostriction induce a substantial electric polarization of 537~$\mu$C~m$^{-2}$ (535~$\mu$C~m$^{-2}$) (without) considering spin-orbit interaction. Te-$5s$ lone electron pair may contribute to the electric polarization, as discussed in Ref.~\cite{ChakrabortyPRB13}. However, since the electric polarization does not correspond to the lowest energy magnetic configuration, we cannot call it a multiferroic material.

\section{\label{sec:conc}Conclusion}
To conclude, we studied bulk hexagonal MnTe within first-principles density functional theory to understand its physical properties, including electronic structure, magnetism, spin-orbit interaction, and ferroelectricity. We begin by examining different exchange-correlation functionals and corrections to find an optimum combination for our calculations. After converging on an appropriate combination of exchange-correlation functional and corrections, we study the system's electronic structure and magnetic properties. Our results reveal an insulating nature of MnTe in its bulk form with a collinear antiferromagnetic order. We estimate the exchange interaction parameters along various exchange paths by mapping the interactions to a spin Hamiltonian. Subsequently, we evaluate the antiferromagnetic ordering temperature based on the exchange interaction strengths using the Metropolis Monte Carlo algorithm to be $\sim$240~K. We find a large spin splitting in the lowest energy antiferromagnetic configuration of the system even without considering spin-orbit interaction, except in the $k_x$-$k_y$ plane where the spin degeneracy remains protected by local symmetry, as long as spin-orbit interaction is not considered. Our systematic and critical examination of spin-orbit interaction via usual band dispersion, 3D band dispersion, isoenergetic contours, and projected spin directions obtained from DFT results compared with an insightful analytical model unequivocally confirms a combined Rashba-Dresselhaus interaction in the $k_x$-$k_y$ plane, around the K point of the system. We attribute the Rashba-Dresselhaus interaction in a structure with centrosymmetric space group $P6_3/mmc$ to a non-centrosymmetric, nonpolar point group corresponding to the Te site and a microscopic electric field along the $(0001)$ direction owing to alternating $+2|-2$ charged layers. A heterostructure involving (0001)-terminated MnTe may host spin splitting only of Rashba-Dresselhaus type, owing effectively to the accessibility of only $k_x$-$k_y$ plane in the reciprocal space for such a heterostructure, provided the magnetic arrangement remains unaltered. If the magnetic arrangement realized in a possible heterostructure is such that the local symmetry protecting antiferromagnetic spin degeneracy is no longer preserved, an antiferromagnetic spin splitting combined with a Rashba-Dresselhaus spin splitting may be realized, with the former dominating over the latter. Finally, we explore the possibility of ferroelectricity in the system. Although the lowest energy antiferromagnetic configuration reveals no electric polarization, another antiferromagnetic configuration with $\uparrow \uparrow \downarrow \downarrow$ arrangement of magnetic moments along the $c$-direction exhibits a pronounced magnetostrictive effect, resulting in substantial electric polarization. Although the bulk material may not have a stable AFM1 configuration to host ferroelectricity, the feature may be present in a possible heterostructure if the relevant magnetic configuration gets stabilized due to the synthesis conditions and/or stress. Further, lattice distortion in the process of formation of a possible heterostructure may stabilize a ferroelectric phase in MnTe. Our studies help understand the electronic structure, magnetism, spin splitting, and spin-orbit interaction in bulk hexagonal MnTe, paving the way for its application in potential spintronic devices.

\begin{acknowledgments}
SR acknowledges CSIR, India, for a research fellowship through grant number 09/1020(0157)/2019-EMR-I. NG acknowledges financial support from SERB, India, through grant number CRG/2021/005320. The use of high-performance computing facilities at IISER Bhopal and PARAM Seva within the framework of the National Supercomputing Mission, India is gratefully acknowledged.
\end{acknowledgments}

%
\end{document}